\documentclass[twocolumn,showpacs,aps,prl,letter,amsmath,amssymb,superscriptaddress]{revtex4-1}
\usepackage{bm,color,bbm}
\usepackage{hyperref, mathtools,graphicx,natbib}

\newcommand{\beq}{\begin{equation}}
\newcommand{\eeq}{\end{equation}}
\newcommand{\bqa}{\begin{eqnarray}}
\newcommand{\eqa}{\end{eqnarray}}
\newcommand{\nn}{\nonumber}

\newcommand{\ket}[1]{ |{#1} \rangle}

\newcommand{\mf}{\mathbf}

\definecolor{maroon}{rgb}{0.7,0,0}

\definecolor{ngreen}{rgb}{0.3,0.7,0.3}

\definecolor{golden}{rgb}{0.8,0.6,0.1}

\graphicspath{ {./figures/} }

\begin{document}
\title{Experimental Verification of Anisotropic Invariance for Three-Qubit States}
\author{Jie Zhu}
\email{These authors contributed equally to this work.}
\affiliation{Laboratory of Quantum Information, University of Science and Technology of China, Hefei 230026}
\affiliation{CAS Center for Excellence in Quantum Information and Quantum Physics, Hefei 230026}

\author{Meng-Jun Hu}
\email{These authors contributed equally to this work.}
\affiliation{Laboratory of Quantum Information, University of Science and Technology of China, Hefei 230026}
\affiliation{CAS Center for Excellence in Quantum Information and Quantum Physics, Hefei 230026}

\author{Shuming Cheng}
\email{shuming.cheng@griffithuni.edu.au}
\affiliation{Centre for Quantum Dynamics, Griffith University, Brisbane, QLD 4111, Australia}
\affiliation{Laboratory of Quantum Information, University of Science and Technology of China, Hefei 230026}
\affiliation{CAS Center for Excellence in Quantum Information and Quantum Physics, Hefei 230026}

\author{Michael J. W. Hall}
\email{michael.hall@griffith.edu.au}
\affiliation{Centre for Quantum Dynamics, Griffith University, Brisbane, QLD 4111, Australia}
\affiliation{Department of Theoretical Physics, Research School of Physics and Engineering, Australian National University, Canberra ACT 0200, Australia}

\author{Chuan-Feng Li}
\affiliation{Laboratory of Quantum Information, University of Science and Technology of China, Hefei 230026}
\affiliation{CAS Center for Excellence in Quantum Information and Quantum Physics, Hefei 230026}

\author{Guang-Can Guo}
\affiliation{Laboratory of Quantum Information, University of Science and Technology of China, Hefei 230026}
\affiliation{CAS Center for Excellence in Quantum Information and Quantum Physics, Hefei 230026}

\author{Yong-Sheng Zhang}
\email{yshzhang@ustc.edu.cn}
\affiliation{Laboratory of Quantum Information, University of Science and Technology of China, Hefei 230026}
\affiliation{CAS Center for Excellence in Quantum Information and Quantum Physics, Hefei 230026}

\begin{abstract}

We experimentally test the recently predicted anisotropic invariance properties of pure three-qubit states, via generation and measurement of polarisation-path entangled three-qubit states. These properties do not require aligned reference frames, and can be determined from measurements on any two of the qubits. They have several applications, such as a universal ordering of pairwise quantum correlations, strong monogamy relations for Bell inequalities and quantum steering, and a complementarity relation for Bell nonlocality versus 3-tangle, some of which we also test. The results indicate that anisotropic invariance, together with the three qubit Bloch vector lengths, can provide a robust and complete set of invariants for such states under local unitary transformations. 

\end{abstract}

\date{\today}

\maketitle

\paragraph{Introduction.---}

Quantum correlations, including entanglement and Bell nonlocality, are considered as useful resources that underlie numerous quantum information processing tasks. When they are employed to accomplish a particular quantum task,  some important issues arise: Does the given quantum state demonstrate certain quantum features? If yes, how useful is this state to accomplish the desired task? The above questions can often be assessed via the invariant quantities that remain unchanged under certain physical operations acting on the state. For example, various resource measures, such as negativity~\cite{VW02} and concurrence of entanglement~\cite{HW97,W98}, have been proposed to study quantum correlations quantitatively, and almost all of them are invariant under local unitary operations. Moreover, well-chosen invariants are also useful for classifying multi-party quantum states qualitatively. For example, the 3-tangle~\cite{CKW00}---an invariant under stochastic local operations and classical communication and under permutations of qubits---determines to which entanglement class a genuinely entangled three-qubit state belongs~\cite{DVC00}.    
Hence, a lot of research works have been devoted to exploring invariants and their applications in the characterisation and quantification of multi-party states. 

Recently, a surprisingly simple invariance property of pure three-qubit states was discovered, with a physical interpretation in terms of the anisotropy of the pairwise spin correlations~\cite{CH17}. This property is captured by two {\it anisotropic} invariants, which can be determined from the spin correlation matrix of any two of the qubits, and which, similarly to the 3-tangle, are invariant under local unitary transformations and under permutations of the qubits. Importantly, these new invariants are particularly well-suited to the experimental measurement, as they are reference-frame independent, i.e., there is  no requirement that the local measuring devices share a common reference frame or alignment. This contrasts with other known invariants for three-qubit states, which typically require full tomography for their evaluation and have no simple physical interpretation~\cite{K99,AAJT01,AACJLT00,S01}. Furthermore, they have a number of useful applications, such as a universal ordering of pairwise quantum correlations for pure three-qubit states, a complementarity relation for Bell nonlocality versus 3-tangle, and strong monogamy relations for various quantum correlations~\cite{CH17}.

In this work, we experimentally verify anisotropic invariance for a set of polarisation-path entangled three-qubit states. Moreover, we use the measured data to investigate entanglement-ordering, Bell nonlocality, and strong monogamy properties of these states.    

\begin{figure*}
	\includegraphics[width=18cm]{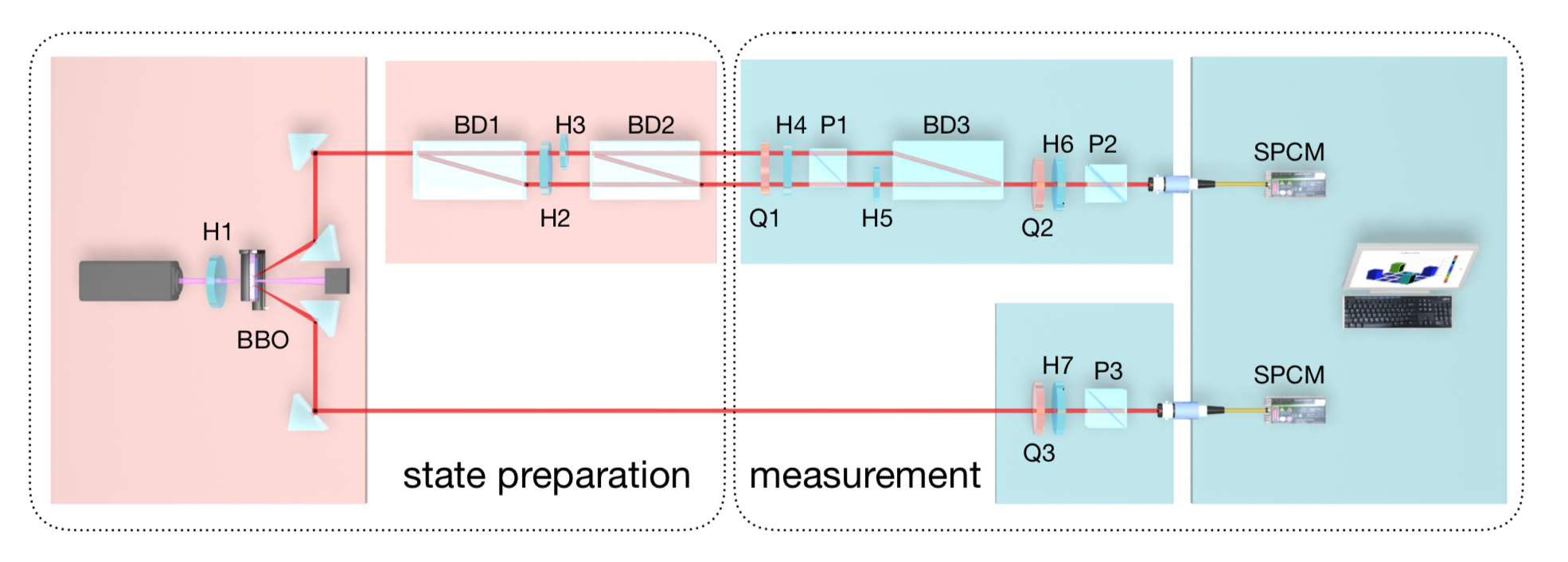}
	\caption{Experimental preparation and measurement of three-qubit states in Eqs.~(\ref{symm}) and (\ref{asymm}). In the state preparation process (pink regions), a pair of entangled photons is first generated via a type-I phase matching nonlinear $\beta$-Barium-Borate crystal. The first photon of each pair then passes through two beam displacers to generate a pair of qubits, encoded in its path- and polarisation degrees of freedom, while the polarisation of the second photon encodes a third qubit. The joint measurement of the qubit spin components is implemented via polarisation analyzers and single photon counting measurements. H1-H7, half-wave plates; Q1-Q7, quarter-wave plates; P1-P3, polarising beam splitters; BD1-BD3, beam displacers; BBO, $\beta$-Barium-Borate crystal; SPCM, single photon counting measurement. } 
	\label{setup}
\end{figure*}

\paragraph{Decomposition of spin correlations.---}

Any two-qubit state $\rho_{AB}$ shared between two parties, Alice and Bob say, is fully characterised by the local Bloch vectors $\bm a=\langle\bm \sigma\otimes I\rangle$ and $\bm b=\langle I\otimes \bm \sigma \rangle$ of the local reduced states, and the $3\times3$ spin correlation matrix $T^{AB}$ with coefficients $T^{AB}_{jk}=\langle\sigma_j\otimes \sigma_k\rangle$. Here $(\sigma_1, \sigma_2, \sigma_3)\equiv \boldsymbol{\sigma}$ are the Pauli spin operators.  The eigenvalues of the symmetric matrix ${\cal S}^{AB}=T^{AB}(T^{AB})^\top$, $s_1\geq s_2\geq s_3\geq 0$, are invariant under local unitary transformations, and are of particular significance in determining the strength of the shared quantum correlations~\cite{HHHH09,MBCPV12,Dakic12,BCPSW14, NV16,TNA16}. 

For example, Alice and Bob can violate the well known Clauser-Horne-Shimony-Holt (CHSH) Bell inequality $\langle\mathcal B_{AB}\rangle^2\leq 4$~\cite{CHSH69}, with $\mathcal B_{AB}(\bm a_1,\bm a_2,\bm b_1,\bm b_2):=\bm \sigma\cdot \bm a_1\otimes \bm \sigma \cdot \bm b_1+\bm \sigma\cdot \bm a_1\otimes \bm \sigma\cdot \bm b_2+\bm \sigma\cdot \bm a_2\otimes \bm \sigma\cdot \bm b_1-\bm \sigma\cdot \bm a_2\otimes \bm \sigma\cdot \bm b_2$, for some choice of measurement directions $\bm a_1,\bm a_2,\bm b_1,\bm b_2$, if and only if ${\mathcal M}^{AB}>1$~\cite{HHH95,BCPSW14}, where
\beq \label{mab}
{\mathcal M}^{AB}:=s^{AB}_1 + s^{AB}_2=\max_{\bm a_1,\bm a_2,\bm b_1,\bm b_2} \frac{1}{4}\mathcal B_{AB}(\bm a_1,\bm a_2,\bm b_1,\bm b_2)^2 . 
\eeq  
This condition also determines whether quantum steering can be witnessed via two spin measurements by each party~\cite{GC16,QZFY17}.

It turns out to be advantageous to consider the natural decomposition of the eigenvalues of ${\cal S}^{AB}$ into isotropic and anisotropic contributions~\cite{CH17},
\beq 
\label{decomp}
s^{AB}_j = s^{AB}_{\rm iso} + \delta s^{AB}_j, \qquad j=1,2,3,
\eeq
where $s^{AB}_{\rm iso}$ denotes the average value $(s_1^{AB} +s_2^{AB} +s_3^{AB})/3$. For isotropic spin corelations, such as for a two-qubit Werner state~\cite{W89}, the anisotropic contributions $\delta s^{AB}_j$ all vanish, yielding $s^{AB}_j = s^{AB}_{\rm iso}$ (with $s^{AB}_j=1$ for a maximally entangled state).  More generally, $s^{AB}_{\rm iso}$ measures the inherent {\it isotropic strength} of the spin correlations, while the $\delta s^{AB}_j$ characterise the inherent {\it anisotropy}. Note that since the $\delta s^{AB}_j$ sum to zero, there are only two independent measures of spin anisotropy. 



\paragraph{Anisotropic invariants for three-qubit states.---} 

For any pure three-qubit state $\ket{\Psi}_{ABC}$ shared by Alice, Bob, and Charlie, there are  reduced density operators $\rho_{AB}$, $\rho_{AC}$,  $\rho_{BC}$ for each pair. The Bloch vectors for Alice, Bob, and Charlie's local states will be denoted by $\mf{a}, \mf{b},$ and $\mf{c}$, respectively, and the corresponding pairwise spin correlation matrices by $T^{AB}$, $T^{AC},$ and $T^{BC}$.

\setlength{\tabcolsep}{2mm}
\begin{table*}
\caption{Experimental results for three-qubit W-class states as per Eq.~(\ref{symm}). The values of the parameters $\phi$ and $\theta$ are specified in the first column. The second column tests the invariance relation~(\ref{isotropy}) for the measured isotropic strengths, while the remaining columns test the anisotropic invariance property~(\ref{aniso1}) for the measured anistropies. Errors are calculated via simulating a Poissonian distribution of photon counts.}
\scalebox{0.9}{
\begin{tabular}{cccccccccccccc} 
\hline
\hline
state & \multicolumn{1}{c}{isotropic strength}&\multicolumn{9}{c}{anisotropies}\\
\hline
$(\phi,\theta)$ 
& $s_{iso}^{AB}+s_{iso}^{AC}+s_{iso}^{BC} $
& $\delta s_{1}^{AB}$ & $\delta s_{1}^{AC}$ & $\delta s_{1}^{BC}$
& $\delta s_{2}^{AB}$ & $\delta s_{2}^{AC}$ & $\delta s_{2}^{BC}$
& $\delta s_{3}^{AB}$ & $\delta s_{3}^{AC}$ & $\delta s_{3}^{BC}$ \\
\hline
$(0,0)$                   & $1.006(2)$   & 0.665(9) & 0.667(4) & 0.667(3)   & -0.332(7) & -0.333(4) & -0.331(4)   & -0.332(7) & -0.333(8) & -0.335(5)\\
$(20^{\circ},0)$          & $0.993(3)$   & 0.396(2) & 0.393(2) & 0.385(3)   & -0.197(4) & -0.190(2) & -0.192(5)   & -0.198(4) &  -0.202(2) & -0.193(1)\\
$(30^{\circ},0)$          & $1.007(6)$   & 0.175(3) & 0.168(4) & 0.165(3)   & -0.104(1) & -0.107(5) & -0.081(8)   & -0.093(3) & -0.098(5) & -0.087(2)\\
$(45^{\circ},0)$          & $1.012(2)$   & 0.003(1) & 0.012(5) & 0.012(2)   & -0.037(3) & -0.036(7) & -0.029(7)   & -0.037(6) & -0.032(5) & -0.041(3)\\
$(30^{\circ},45^{\circ})$ & $1.004(2)$   & 0.127(3) & 0.110(2) & 0.116(2)   & -0.056(5) & -0.044(2) & -0.056(3)   & -0.071(2) & -0.066(2) & -0.063(3)\\
$(45^{\circ},45^{\circ})$ & $0.981(9)$   & 0.095(4) & 0.112(6) & 0.109(6)   & 0.071(4)  & 0.075(5)  & 0.076(4)    & -0.166(3) & -0.187(4) & -0.175(5)\\
$(30^{\circ},30^{\circ})$ & $1.010(9)$   & 0.139(4) & 0.136(5) & 0.131(5)   & -0.062(9) & -0.059(1) & -0.063(8)   & -0.075(8) & -0.076(7) & -0.067(1)\\
$(45^{\circ},30^{\circ})$ & $1.007(7)$   & 0.074(2) & 0.065(4) & 0.066(6)   & -0.053(3) & -0.052(3) & -0.059(8)   & -0.128(3) & -0.117(2) & -0.126(4)\\
$(45^{\circ},15^{\circ})$ & $1.002(7)$   & 0.024(1) & 0.033(1) & 0.025(7)   & 0.021(5)  & 0.015(7)  & 0.013(7)    & -0.045(6) & -0.048(8) & -0.039(4)  \\
\hline
\end{tabular}}
\label{tab1}
\end{table*}

The corresponding isotropic strengths of the three two-qubit states are easily shown to satisfy~\cite{CH17}
\beq
s_{\rm iso}^{AB}+s_{\rm iso}^{AC}+s_{\rm iso}^{BC}=1. 
\label{isotropy}
\eeq
Thus, the sum of the isotropic strengths is a constant, although the individual strengths can vary from pair to pair. In contrast, the spin anisotropy properties are {\it identical} for each pair of qubits~\cite{CH17}: 
\begin{align}
\delta s^{AB}_j=\delta s^{AC}_j=\delta s^{BC}_j,\qquad j=1,2,3. \label{aniso1} 
\end{align} 
This invariance property is highly nontrivial, and immediately indicates that the pairwise anisotropies are not only invariant under local unitary transformations, but also under any permutation of the qubits or parties. In particular, they can be determined quantitatively from measurements on any pair of qubits, and without any need for a shared reference frame. Moreover, they are independent of the 3-tangle, and can be generalised  to mixed states via a convex-roof construction~\cite{CH17}. The distribution of their values over the space of pure three-qubit states is studied in~\cite{Z18}. 

The anisotropic invariance property in Eq.~(\ref{aniso1}) has a number of useful applications, including a strong monogamy relation for the Bell nonlocality of general three-qubit states, and a universal ordering of correlations for pure three-qubit states~\cite{CH17}. In addition to experimentally testing invariance properties~(\ref{isotropy}) and~(\ref{aniso1}), we experimentally investigate these applications below.  

\paragraph{Experimental setup.---}
A 140-mW laser with a wavelength of 404 nm is used to pump a combined type-I phase matching $\beta$-barium borate (BBO) crystal to generate a pair of degenerate photons via the spontaneous parametric down-conversion (SPDC) process. The state of the photon pair is described by $\cos{(\phi)}\left| HH \right\rangle+\sin{ (\phi)}\left| VV \right\rangle$~\cite{KWWAE99}, where $|H\rangle$ represents horizontal polarisation and $|V\rangle$ represents vertical polarisation, and the phase $\phi$ is determined by the rotation angle of a half-wave plate (HWP) before the BBO crystal. The polarisation degrees of freedom of the photon pair encode two qubits, labeled as $A$ and $C$ respectively. As shown in Fig.~\ref{setup}, the first photon is then sent through two beam displacers (BD1, BD2) sandwiching two HWPs (H2, H3), to entangle its path- and polarisation- degrees of freedom, while the second photon goes directly to the measurement process.  In particular, BD1 in the upper (pink) box acts such that the vertically-polarised component of the first photon passes into the upper path, while the horizontally-polarised component passes into the lower path. This path degree of freedom of the first photon encodes a qubit labeled as $B$. Thus, the state preparation process generates entangled three-qubit states.

In the experiment, we prepared two classes of pure three-qubit states for testing anisotropic invariance. The first class of states generated is of the form 

\begin{align}\nn
\left| \psi \right\rangle_{ABC}=&\cos{(\phi)}\left| 1\right\rangle \left| 1 \right\rangle \left| 0\right\rangle +\sin{(\phi)} \cos{(\theta)}\left| 0\right\rangle\left| 1 \right\rangle \left|1\right\rangle \\ & + \sin{(\phi)}\sin{(\theta)}\left| 1\right\rangle\left| 0 \right\rangle \left|1\right\rangle,
\label{symm}
\end{align}
where the angle $\theta/2$ is determined via rotating the HWP (H3) placed in the upper path. Here each horizontal polarisation state $\left|H\right\rangle$ and the upper path state are  encoded by qubit state $|0\rangle$, while the vertical polarisation state $\left|V\right\rangle$ and the lower path  state are encoded by qubit state $|1\rangle$.  Note that states of the form of Eq.~(\ref{symm}) are  W-class states, and hence have zero 3-tangle~\cite{DVC00}. In addition to the class of states as per Eq.~(\ref{symm}), we also prepared a class of three-qubit states of the form

\begin{align}\nn
\left| \chi \right\rangle_{ABC}=&\cos{(\phi')}\left| 1\right\rangle \left| 1 \right\rangle \left| 0\right\rangle +\frac{1}{\sqrt{2}}\sin{(\phi')} \left| 0\right\rangle\left| 1 \right\rangle \left|1\right\rangle +\\ &\frac{1}{\sqrt{2}}\sin{(\phi')}\left| 0\right\rangle\left| 0 \right\rangle \left|1\right\rangle,
\label{asymm}
\end{align}
 via inserting a $45^{\circ}$-HWP after BD2 in the upper path.  Here $\phi'= 90^{\circ}-\phi$.  These are Greenberg-Horne-Zeilinger (GHZ)-class states, with  nonzero 3-tangles~\cite{DVC00}.

Joint measurements of the three spin components ${\sigma}^{A}_j,{\sigma}^{B}_k,\sigma^{C}_l$ were implemented on the three qubits, for each possible choice $j,k,l=1,2,3$, as depicted in the right half of Fig.~\ref{setup} (blue region).  The measurement process includes three polarisation analyzers in the configuration, each of which consists of a HWP, a quarter-wave plate, and a polarising beam splitter, to obtain the state information. We measured the path information corresponding to qubit B via conversion from path to polarization states. If a photon is in the upper path after BD2, following the preparation process, then its polarization is horizontal when it arrives at Q2; while if it is in the down path it  will  be  vertically-polarized.   Therefore,  the polarisation analyzer (Q2, H6, P2) obtains the path state information via polarization state tomography.

Finally,  single photon counting measurements (SPCMs) were used to collect the experimental data, as relative frequencies corresponding to the joint probability distribution $\textsl{P}_{jkl}(\alpha,\beta,\gamma)$, where $\alpha,\beta,\gamma=\pm1$ label the respective outcomes for  ${\sigma}^{A}_j,{\sigma}^{B}_k,\sigma^{C}_l$. 
The pairwise spin correlation matrix $T^{AB}$ for qubits $A$ and $B$ is  experimentally determined via
\begin{equation}
T^{AB}_{jk} = \langle {\sigma}^{A}_j\otimes {\sigma}^{B}_k \rangle=\sum_{\alpha,\beta,\gamma=\pm1} \alpha \beta\, \textsl{P}_{jkl}(\alpha,\beta,\gamma) 
\end{equation}
(for any choice of $l$), with similar expressions for $T^{AC}$ and $T^{BC}$. The isotropic strengths $s_{\rm iso}^{AB},s_{\rm iso}^{AC},s_{\rm iso}^{BC}$ and anisotropies $\delta s^{AB}_j,\delta s^{AC}_j,\delta s^{BC}_j$ can then be determined via the eigenvalues of the corresponding matrices ${\cal S}^{AB}, {\cal S}^{AC}, {\cal S}^{BC}$, as per Eq.~(\ref{decomp}).

\setlength{\tabcolsep}{2.8mm}
\begin{table}[!h]
\renewcommand\arraystretch{1.5}
	\caption{Five typical three-qubit W-class states, as per Eq.~(\ref{symm}), are choosen to test the strong monogamy relation~(\ref{bell}) for Bell nonlocality, via the corresponding Horodecki parameters (columns 2 and 3), and the maximal violation of the CHSH-Bell inequality (columns 4 and 5). Errors are calculated via simulating Poissonian distribution of photon counts.}
    \scalebox{0.86}{
	\begin{tabular}{ccccc}
		\hline
		\hline
		$(\phi,\theta)$  & ${\cal M}^{AB}$ & ${\cal M}^{AC}$ & $\langle {\mathcal B}_{AB}\rangle_{\rm max}^2/4$ & $\langle {\mathcal B}_{AC}\rangle_{\rm max}^2/4$\\
		\hline
		$(30^{\circ},45^{\circ})$  & 0.323(3)  & 0.931(2)  & 0.257(3) & 0.925(3) \\
		$(45^{\circ},45^{\circ})$  & 0.499(2)  & 1.014(6)  & 0.493(9) & 0.988(4) \\
		$(30^{\circ},30^{\circ})$  & 0.310(9)  & 1.334(4)  & 0.257(6) & 1.300(2) \\
		$(45^{\circ},30^{\circ})$  & 0.384(4)  & 1.500(3)  & 0.367(2) & 1.485(6) \\
		$(45^{\circ},15^{\circ})$  & 0.136(7)  & 1.864(6)  & 0.119(4) & 1.857(2) \\
		\hline
	\end{tabular}}
	\label{tab2}
\end{table}

\paragraph{Results---}
In this experiment, we prepared twelve states in total: nine W-class states as per Eq.~(\ref{symm}), and three GHZ-class states as per Eq.~(\ref{asymm}). The efficiency of the SPCM was about $68\%$; the coincident window was 3 ns and the coincident count rate was about 1000 ${\rm s}^{-1}$. We also performed quantum state tomography on our prepared states~\cite{JKMW01}, and achieved an average fidelity of $99.26\%\pm0.3\%$. The statistical errors are evaluated via the simulation of the Poissonian distribution of photon counts. There are also alignment errors, including an uncertainty of about $0.5^{\circ}$ for the HWP settings. Our results for W-class states are listed in Tables~\ref{tab1}--\ref{tab3}, while those for  GHZ-class states are given in the Appendix.

\subparagraph{Invariants---} First, it is seen from Table~\ref{tab1} that the sum of the isotropic strengths is unity to within error, thus verifying Eq.~(\ref{isotropy}) for pure three-qubit states. The anisotropies $\delta s^{AB}_{j}, \delta s^{AC}_{j}, \delta s^{BC}_{j}$ for $j=1,2,3$ are similarly found to be identical, to within experimental error, as predicted by anisotropic invariance as per Eq.~(\ref{aniso1}).

\subparagraph{Strong monogamy---} It follows from invariance properties~(\ref{isotropy}) and~(\ref{aniso1}) that the Horodecki parameter defined in Eq.~(\ref{mab}) can be rewritten as ${\mathcal M}^{AB}= 1+s_3^{AB}-s_3^{AC}-s^{BC}_3$ for pure three-qubit states~\cite{CH17}. A similar expression for ${\mathcal M}^{AC}$ is obtained by permuting the labels, yielding~\cite{CH17,CL18}
\beq \label{monog}
{\mathcal M}^{AB}+{\mathcal M}^{AC}=2(1-s_3^{BC})\leq 2.
\eeq
This immediately implies the strong monogamy relation  
\beq
\langle\mathcal B_{AB}(\bm a_1,\bm a_2,\bm b_1,\bm b_2)\rangle^2 + \langle\mathcal B_{AC}(\bm a'_1,\bm a'_2,\bm c_1,\bm c_2)\rangle^2 \leq 8
\label{bell}
\eeq
for the CHSH Bell inequality, via Eq.~(\ref{mab}), for all three-qbuit states and all choices of the measurement directions. This is stronger than previous monogamy relations, which either give weaker restrictions on the expectation values~\cite{SG01,SGPRA01} or require fixed measurement directions $\bm a_1=\bm a'_1$, $\bm a_2=\bm a'_2$ for Alice~\cite{TV06, BCPSW14}.

The measured values of the Horodecki parameters ${\cal M}^{AB}, {\cal M}^{AC}$, given in Table~\ref{tab2} for several W-class states, indicate that Alice and Charlie can violate the CHSH Bell inequality via suitable measurements for the states corresponding to $(\phi,\theta)=(30^{\circ},30^{\circ}), (45^{\circ},30^{\circ}), (45^{\circ},15^{\circ})$ in Eq.~(\ref{symm}), whereas Alice and Bob cannot. This exemplifies the strong monogamy relation~(\ref{monog}), which may also be quantitatively checked via summing the Horodecki parameters in the second two columns of Table~\ref{tab2}.



\setlength{\tabcolsep}{2mm}
\begin{table}[!h]
\renewcommand\arraystretch{1.4}
	\caption{Relative degrees of Bell nonlocality, the Horodecki parameter, isotropic strength, and Bloch vector lengths, for states as per Eq.~(\ref{symm}), to test the universal ordering predicted in Eq.~(\ref{ab}). Errors are calculated via simulating Poissonian distribution of photon counts.}
	\scalebox{0.75}{
		\begin{tabular}{ccccc}
			\hline
			\hline
			$(\phi,\theta)$           & $({\mathcal C}^{AB})^2-({\mathcal C}^{AC})^2$ & $({\cal M}^{AB}-{\cal M}^{AC})/2 $ & $s_{iso}^{AB}-s_{iso}^{AC}$   & $|\bm c|^2-|\bm b|^2 $\\
			\hline
			$(0,0)$                   &  -2.783e-04       &  8.000e-04        & -0.002(1)        & -7.960e-04 \\
			$(20^{\circ},0)$          &  -0.412(4)        &  -0.405(3)        & -0.403(3)        & -0.410(5) \\
			$(30^{\circ},0)$          &  -0.745(6)        &  -0.735(4)        & -0.738(8)        & -0.746(8) \\
			$(45^{\circ},0)$          &  -0.985(3)        &  -0.977(7)        & -0.980(2)        & -0.995(6)\\
		    $(30^{\circ},45^{\circ})$ &  -0.309(7)        &  -0.304(4)        & -0.306(2)        & -0.309(4)\\
			$(45^{\circ},45^{\circ})$ &  -0.237(7)        &  -0.257(5)        & -0.247(5)        & -0.251(2)\\
			$(30^{\circ},30^{\circ})$ &  -0.502(5)        &  -0.512(2)        & -0.512(2)        & -0.517(3)\\
			$(45^{\circ},30^{\circ})$ &  -0.543(6)        &  -0.558(2)        & -0.563(5)        & -0.558(3)\\
			$(45^{\circ},15^{\circ})$ &  -0.861(2)        &  -0.862(4)        & -0.870(5)        & -0.867(9)\\
			\hline
	\end{tabular}}
	\label{tab3}
\end{table}

Further, to directly check that the strong monogamy relation~(\ref{bell}) is satisfied,  we also  experimentally estimated the maximum value of the corresponding CHSH parameters for the states in Table~\ref{tab2}. In particular, we used the theoretically optimal measurement directions for these states to measure the corresponding maximum values $\langle {\mathcal B}_{AB}\rangle_{\rm max}$ and $\langle {\mathcal B}_{AC}\rangle_{\rm max}$ of $\langle\mathcal B_{AB}(\bm a_1,\bm a_2,\bm b_1,\bm b_2)\rangle$ and $\langle\mathcal B_{AC}(\bm a'_1,\bm a'_2,\bm c_1,\bm c_2)\rangle$. For example, the state $(45^{\circ},45^{\circ})$ is measured using the directions $\bm a_1=\bm a'_1=(1,0,0)$, $\bm a_2=\bm a'_2=(0,1,0)$, $\bm b_1=\bm c_1=\frac{1}{\sqrt{2}}(1,1,0)$ and $\bm b_2=\bm c_2=\frac{1}{\sqrt{2}}(1,-1,0)$

 The results are shown in the last two columns of Table~\ref{tab2}, and may be compared to the corresponding values of ${\cal M}^{AB}$ and ${\cal M}^{AC})$. In particular, Eq.~(\ref{bell}) is verified to within experimental error, taking into account the alignment errors of $\pm 0.5^\circ$ for each HWP. 

\subparagraph{Ordering of correlations---} The final application of aniostropic invariance which we test here is a universal ordering of pairwise correlations such as entanglement and Bell nonlocality. For example, from invariance properties~(\ref{isotropy}) and~(\ref{aniso1}) one finds that ${\mathcal M}^{AB}\geq {\mathcal M}^{AC}$ if and only if (iff) ${\mathcal C}^{AB}\geq {\mathcal C}^{AC}$ iff  $s_{\rm iso}^{AB}\geq s_{\rm iso}^{AC}$~\cite{CH17}, where ${\mathcal C}^{AB}$ denotes the concurrence of $\rho_{AB}$~\cite{HW97,W98}. Indeed, this ordering
of pairwise correlation strengths may be strengthened to the quantitative relation~\cite{CH17}
\begin{align} 
({\mathcal C}^{AB})^2- ({\mathcal C}^{AC})^2=&\frac{ {\mathcal M}^{AB}-{\mathcal M}^{AC}}{2}
=  s_{\rm iso}^{AB}-s_{\rm iso}^{AC}\nn \\ &~=|\mf{c}|^2-|\mf{b}|^2 \label{ab},
\end{align}
and its permutations. 
The experimental results in Table~\ref{tab3} verify this ordering relation of pairwise correlations as predicted in Eq.~(\ref{ab}). The data for concurrence and Bloch vector length is calculated via state tomography."

	

\paragraph{Conclusion.---} We have experimentally verified the anisotropic invariance properties of pure three-qubit states, as well as their connections to strong monogamy relations, Bell inequalities and the ordering of pairwise correlations. The anisotropic invariants in Eq.~(\ref{aniso1}) are found to be experimentally robust to minor imperfections. Together with the Bloch vector lengths $|\bm a|, |\bm b|, |\bm c|$, this robustness makes them useful as an experimental tool for characterizing the properties of such states up to local unitary invariance~\cite{CH17}, particularly since there is no need for a common reference frame and only pairwise measurements are required. It would also be of interest to test the statistical secret sharing scheme proposed in~\cite{CH17}, based on the anisotropic invariance property.

\paragraph{Acknowledgements.---} We thank Xiaomin Hu, Yongnan Sun, Yu Guo and Zhaodi Liu for helpful discussions. Part of this work was done during a visit to Griffith Unversity by Meng-Jun Hu. This research was supported by the National Natural Science Foundation of China (Grants Nos. 11674306 and 61590932), National Key R$\&$D Program (No. 2016YFA0301300 and No. 2016YFA0301700), and the Anhui Initiative in Quantum Information Technologies.

\begin{thebibliography}{42}

\bibitem{VW02}
G. Vidal and R. F. Werner, Computable measure of entanglement, \href{https://journals.aps.org/pra/abstract/10.1103/PhysRevA.65.032314}{Phys. Rev. A \textbf{65}, 032314 (2002).}

\bibitem{HW97} 
S. Hill and W. K. Wootters, Entanglement of a Pair of Quantum Bits, 
\href{https://journals.aps.org/prl/abstract/10.1103/PhysRevLett.78.5022}{Phys. Rev. Lett. \textbf{78}, 5022 (1997). }

\bibitem{W98} 
W. K. Wootters, Entanglement of Formation of an Arbitrary State of Two Qubits, 
\href{https://journals.aps.org/prl/abstract/10.1103/PhysRevLett.80.2245}{Phys. Rev. Lett. \textbf{80}, 2245 (1998).} 

\bibitem{CKW00}
V. Coffman, J. Kundu, and W. K. Wootters, Distributed entanglement, \href{https://journals.aps.org/pra/abstract/10.1103/PhysRevA.61.052306}{Phys. Rev. A \textbf{61}, 052306 (2000).}
	
\bibitem{DVC00}  W. D\"ur, G. Vidal, and J. I. Cirac, Three qubits can be entangled in two inequivalent ways,
\href{https://journals.aps.org/pra/abstract/10.1103/PhysRevA.62.062314}{Phys. Rev. A \textbf{62}, 062314 (2000).} 

\bibitem{CH17}
S. Cheng  and M. J. W. Hall, Anisotropic invariance and the distribution of quantum correlations, \href{https://journals.aps.org/prl/abstract/10.1103/PhysRevLett.118.010401}{Phys. Rev. Lett. \textbf{118}, 010401 (2017). }

\bibitem{S01}
A. Sudbery, On local invariants of pure three-qubit states, \href{http://iopscience.iop.org/article/10.1088/0305-4470/34/3/323/meta}{J. Phys. A. Math. Gen. \textbf{34}, 643 (2001).}

\bibitem{AAJT01}
A. Ac\'{\i}n, A. Andrianov, E. Jan\'e, and R. Tarrach, Three-qubit pure-state canonical forms, \href{http://iopscience.iop.org/article/10.1088/0305-4470/34/35/301/meta}{J. Phys. A: Math. Gen. \textbf{34},
6725 (2001).}

\bibitem{AACJLT00}
A. Ac\'{\i}n, A. Andrianov, L. Costa, E. Jan\'e, J. I. Latorre, and R. Tarrach, Generalized Schmidt Decomposition and Classification of Three-Quantum-Bit States, \href{https://journals.aps.org/prl/abstract/10.1103/PhysRevLett.85.1560}{Phys. Rev. Lett. \textbf{85}, 1560 (2000).}

\bibitem{K99}
J. Kempe, Multiparticle entanglement and its applications to cryptography, \href{https://journals.aps.org/pra/abstract/10.1103/PhysRevA.60.910}{Phys. Rev. A \textbf{60}, 910 (1999). }

\bibitem{HHHH09}
R. Horodecki, P. Horodecki, M. Horodecki, and K. Horodecki, Quantum entanglement, \href{https://journals.aps.org/rmp/abstract/10.1103/RevModPhys.81.865}{Rev. Mod. Phys. \textbf{81}, 865 (2009).}

\bibitem{MBCPV12}
K. Modi, A. Brodutch, H. Cable, T. Paterek, and V. Vedral, The classical-quantum boundary for correlations: Discord and related measures, \href{https://journals.aps.org/rmp/abstract/10.1103/RevModPhys.84.1655}{Rev. Mod. Phys. \textbf{84}, 1655 (2012).}

\bibitem{Dakic12}  B. Daki\'c, Y. O. Lipp, X. Ma, M. Ringbauer, S. Kropatschek, S. Barz, T. Paterek, V. Vedral, A. Zeilinger, \'C. Brukner, and P. Walther, Quantum discord as resource for remote state preparation,
\href{https://www.nature.com/articles/nphys2377}{Nat. Phys. \textbf{8}, 666 (2012).}  

\bibitem{BCPSW14}
N. Brunner, D. Cavalcanti, S. Pironio, V. Scarani, and S. Wehner, Bell nonlocality, \href{https://journals.aps.org/rmp/abstract/10.1103/RevModPhys.86.419}{Rev. Mod. Phys. \textbf{86}, 419 (2014).}

\bibitem{NV16} 
H. Chau Nguyen and T. Vu, Necessary and sufficient condition for steerability of two-qubit states by the geometry of steering outcomes, \href{http://iopscience.iop.org/article/10.1209/0295-5075/115/10003/meta#}{Europhys. Lett. \textbf{115}, 10003 (2016).}

\bibitem{TNA16} M. M. Taddei, R. V. Nery, and L. Aolita, Necessary and sufficient conditions for multipartite Bell violations with only one trusted device
\href{https://journals.aps.org/pra/abstract/10.1103/PhysRevA.94.032106}{Phys. Rev. A \textbf{94}, 032106 (2016).} 

\bibitem{CHSH69}
J. F. Clauser, M. A. Horne, A. Shimony, and R. A. Holt, Proposed Experiment to Test Local Hidden-Variable Theories, \href{https://journals.aps.org/prl/abstract/10.1103/PhysRevLett.23.880}{Phys. Rev. Lett. \textbf{23}, 880 (1969).}

\bibitem{HHH95}
R. Horodecki, P. Horodecki, and M. Horodecki, Violating Bell inequality by mixed spin-$1/2$ states: necessary and sufficient condition, \href{https://www.sciencedirect.com/science/article/pii/037596019500214N}{Phys. Lett. A \textbf{200}, 340 (1995).}

\bibitem{GC16}
P. Girdhar and E. G. Cavalcanti, All two-qubit states that are steerable via Clauser-Horne-Shimony-Holt-type correlations are Bell nonlocal,  \href{https://journals.aps.org/pra/abstract/10.1103/PhysRevA.94.032317}{Phys. Rev. A \textbf{94}, 032317 (2016).}

\bibitem{QZFY17}
Q. Quan, H. Zhu, H. Fan, and W. Yang, Einstein-Podolsky-Rosen correlations and Bell correlations in the simplest scenario, \href{https://journals.aps.org/pra/abstract/10.1103/PhysRevA.95.062111}{Phys. Rev. A \textbf{95}, 062111 (2017). }


\bibitem{W89}
R. F. Werner, Quantum states with Einstein-Podolsky-Rosen correlations admitting a hidden-variable model, \href{https://journals.aps.org/pra/abstract/10.1103/PhysRevA.40.4277}{Phys. Rev. A \textbf{40}, 4277 (1989).}



\bibitem{Z18} M. Enr\'iquez, F. Delgado,  and K. \.{Z}yczkowski, Entanglement of three-qubit random pure states, \href{https://www.mdpi.com/1099-4300/20/10/745}{Entropy \textbf{20}, 745 (2018).}


\bibitem{CL18}
S. Cheng and L. Liu, Monogamy relations of nonclassical correlations for multi-qubit states,  \href{https://www.sciencedirect.com/science/article/pii/S0375960118304511}{Phys. Lett. A \textbf{382} (26), 1716 (2018).}

\bibitem{SG01}
V. Scarani and N. Gisin, Quantum Communication between $N$ Partners and Bell's Inequalities, \href{https://journals.aps.org/prl/abstract/10.1103/PhysRevLett.87.117901}{Phys. Rev. Lett. \textbf{87}, 117901 (2001).}

\bibitem{SGPRA01}
V. Scarani and N. Gisin, Quantum key distribution between $N$ partners: Optimal eavesdropping and Bell’s inequalities, \href{https://journals.aps.org/pra/abstract/10.1103/PhysRevA.65.012311}{Phys. Rev. A \textbf{65}, 012311  (2001).}

\bibitem{TV06}
B. Toner and F. Verstraete, Monogamy of Bell correlations and Tsirelson's bound, \href{https://arxiv.org/abs/quant-ph/0611001}{arXiv:quant-ph/0611001.}


\bibitem{KWWAE99}
P. G. Kwiat, E. Waks, A. G. White, I. Appelbaum, and P. H. Eberhard, Ultrabright source of polarization-entangled photons, \href{https://journals.aps.org/pra/abstract/10.1103/PhysRevA.60.R773}{Phys. Rev. A \textbf{60}, R773 (1999).}


\bibitem{JKMW01}
D. F. V. James, P. G. Kwiat, W. J. Munro and A. G. White, Measurement of qubits, \href{https://journals.aps.org/pra/abstract/10.1103/PhysRevA.64.052312}{Phys. Rev. A \textbf{64}, 052312 (2001).}



\end{thebibliography}

%

\newpage
\onecolumngrid

\section{Appendix}

\subsection{Fidelity of states}
As noted in the main text, we tomographically reconstructed the density matrices, and calculated their fidelities with the corresponding W-class states in Table~I.  The results are shown in Table~\ref{tab4} below. 

\linespread{1.5}
\setlength{\tabcolsep}{3.5mm}
\begin{table*}[!hbp]
	\caption{Measured fidelities }
\begin{center}
\begin{tabular}{c | c c c c c}

\hline
\hline
$(\phi,\theta)$  &  $(0,0)$   &  $(20^{\circ},0)$ & $(30^{\circ},0)$ & $(45^{\circ},0)$ & $(30^{\circ},45^{\circ})$    \\
\hline
Fidelity         &  $99.34\%$ &  $99.79\%$        & $98.76\%$        & $99.30\%$        & $99.55\%$   \\ 
\hline
\hline
$(\phi,\theta)$ & $(45^{\circ},45^{\circ})$ & $(30^{\circ},30^{\circ})$ & $(45^{\circ},30^{\circ})$ & $(45^{\circ},15^{\circ})$ & average   \\
\hline
Fidelity        & $98.88\%$                 & $99.09\%$                 & $99.15\%$                 &  $99.51\%$            & $99.26\%$  \\ 
\hline
\end{tabular}
\end{center}
\label{tab4}
\end{table*}

\subsection{Results for GHZ-class states}
As noted in the main text, we also tested anisotropic invariance and its applications for three GHZ-class states, of the form 
\begin{equation}
\left| \psi \right\rangle=\cos{(\phi')}\left| 1\right\rangle \left| 1 \right\rangle \left| 0\right\rangle +\frac{1}{\sqrt{2}}\sin{(\phi')} \left| 0\right\rangle\left| 1 \right\rangle \left|1\right\rangle +\frac{1}{\sqrt{2}}\sin{(\phi')}\left| 0\right\rangle\left| 0 \right\rangle \left|1\right\rangle
\label{asymmapp}
\end{equation}
(corresponding to Eq.~(9) of the main text).

Tables~\ref{tab5} and~\ref{tab6} give experimental results for several of the same quantities given in Tables~I-III of the main text.  These results are, within experimental error, in full accord with predictions.

\linespread{1.5}
\setlength{\tabcolsep}{3.5mm}
\begin{table}[!hbp]
\caption{Experimental results for GHZ-class states as per Eq.~(\ref{asymmapp}).}
\scalebox{0.85}{
\begin{tabular}{ccccccccccc}
\hline
\hline
 state & \multicolumn{1}{c}{isotropic strength}&\multicolumn{9}{c}{anisotropies}\\
 \hline
$\phi'$ 
&  $s^{AB}_{iso}+s^{AC}_{iso}+s^{BC}_{iso}$  
& $\delta s_{1}^{AB}$ & $\delta s_{1}^{AC}$ & $\delta s_{1}^{BC}$ 
& $\delta s_{2}^{AB}$ & $\delta s_{2}^{AC}$ & $\delta s_{2}^{BC}$ 
& $\delta s_{3}^{AB}$ & $\delta s_{3}^{AC}$ & $\delta s_{3}^{BC}$ \\

\hline
$20^{\circ}$ & 0.995(6) & 0.525(3) & 0.525(4) & 0.524(8) & -0.262(5) & -0.259(9) & -0.262(2) & -0.262(6) & -0.264(5) &-0.262(1)\\
$30^{\circ}$ & 0.992(6) & 0.413(4) & 0.413(3) & 0.415(4) & -0.206(7) & -0.201(9) & -0.207(5) & -0.206(7) & -0.210(7) &-0.207(5)\\
$45^{\circ}$ & 0.996(8) & 0.333(4) & 0.331(4) & 0.334(3) & -0.202(5) & -0.166(3) & -0.156(6) & -0.166(8) & -0.166(9) &-0.174(1)\\
\hline
\end{tabular}}
\label{tab5}
\end{table}

\linespread{1.5}
\setlength{\tabcolsep}{3.5mm}
\begin{table}[!hbp]
\caption{Relative degrees of Bell nonlocality, isotropic strength, and Bloch vector lengths, for GHZ-class states as per Eq.~(\ref{asymmapp}).  }
\begin{tabular}{cccccc}
\hline
\hline
$\phi'$       & $({\mathcal C}^{AB})^2-({\mathcal C}^{AB})^2$ & $\frac{1}{2}({\cal M}^{AB}-{\cal M}^{AC})$ & $s_{iso}^{AB}-s_{iso}^{AC}$ & $c^2-b^2$ \\
\hline
$20^{\circ} $ &  -0.199(5)                                     & -0.209(8) & -0.208(1)  & -0.199(6) \\
$30^{\circ} $ &  -0.372(1)                                     &  -0.373(3) & -0.371(7)  & -0.367(5) \\
$45^{\circ} $ &  -0.456(6)                                     & -0.499(7)  & -0.495(2) & -0.480(3) \\
\hline
\\
\end{tabular}
\label{tab6}
\end{table}

\end{document}